\documentclass[12pt]{article}

\newcommand {\eqref} [1] {(\ref {#1})}
\newcommand {\beq} {\begin{equation}} 
\newcommand {\eeq} {\end{equation}}
 \newcommand {\ber}{\begin{eqnarray*}}
 \newcommand {\eer} {\end{eqnarray*}}
\newcommand {\bea}{\begin{eqnarray}}
 \newcommand {\eea} {\end{eqnarray}} 
\newcommand{\non}{\nonumber}

\def\a{\alpha}          
           
\def\c{\gamma}  \def\C{\Gamma}  
\def\d{\delta}

\def\l{\lambda}

\def\th{\theta}

\def\cN{{\cal N}}

\def\rr{{\rm r}}
\def\rv{{\rm v}}

\newcommand{\be}{\begin{equation}}
\newcommand{\ee}{\end{equation}}
\newcommand{\eq}[1]{(\ref{#1})}
\let\bm=\bibitem

\begin{document}
\hfill{hep-th/0208148}

\hfill{LPTENS-02/41}

\hfill{PAR-LPTHE 02-46}

\vspace{20pt}

\begin{center}

{\LARGE \bf A note on string interaction on the pp-wave background}
\vspace{30pt}
 
{\bf Chong-Sun Chu$^{a,}$\footnote{on leave of absence from 
University of Durham, UK}, 
Valentin V. Khoze$^{b}$,
Michela Petrini$^{c}$, \\Rodolfo Russo$^{d}$,
Alessandro Tanzini$^{e}$
}

\vspace{15pt}
{\small \em 
\begin{itemize}
\item[$^a$]
Department of Physics, National Tsing Hua University, 
Hsinchu, Taiwan 300, R.O.C.
\item[$^b$]
Centre for Particle Theory and IPPP,
University of Durham, Durham, DH1 3LE, UK
\item[$^c$] 
Centre de Physique Th{\'e}orique, Ecole
Polytechnique\footnote{Unit{\'e} mixte du CNRS et de l'EP, UMR
7644}, 91128 Palaiseau Cedex, France
\item[$^d$]
Laboratoire de Physique Th\'eorique de
l'Ecole Normale Sup\'erieure,  \\
24 rue Lhomond, {}F-75231 Paris Cedex 05, France 
\item[$^e$]
LPTHE, Universit\'e de Paris~VI-VII, 4 Place Jussieu 75252 Paris Cedex 05,
France, \\ 
and INFN, Roma, Italy.
\end{itemize}
}

\vskip .1in {\small \sffamily chong-sun.chu@durham.ac.uk, 
valya.khoze@durham.ac.uk,petrini@cpht.polytechnique.fr,
\\rodolfo.russo@lpt.ens.fr,tanzini@lpthe.jussieu.fr}
\vspace{50pt}

{\bf Abstract}
\end{center}

We consider type IIB string interaction on the maximally
supersymmetric pp-wave background and discuss how the bosonic
symmetries of the background are realized. This analysis shows that
there are some interesting differences with respect to the flat--space
case and suggests modifications to the existing form of the string
vertex. We focus on the zero--mode part which is responsible for some
puzzling string predictions about the $\cN=4$ SYM side. We show that
these puzzles disappear when a symmetry preserving string interaction
is used.

\setcounter{page}0
\setcounter{footnote}0
\thispagestyle{empty}
\newpage

\section{Introduction}

The duality~\cite{Maldacena:1997re} between IIB string theory on
$AdS_5 \times S_5$ and ${\cal N} = 4$ Super Yang--Mills theory
realized, for the first time, the idea that the planar limit of a
quantum gauge theory is equivalent to a classical string
theory. However, string theory on the $AdS_5 \times S_5$ background
remains still largely intractable and even the free spectrum for this
case is not known. Thus, most of the checks of the above duality have
been done in the supergravity limit ($\alpha' \to 0$), where the gauge
theory is strongly coupled ($\lambda = g_{YM}^2 N \to \infty$).
In~\cite{Gubser:1998bc} a concrete rule was given for comparing, in
the $\lambda \to\infty$ limit, dynamical quantities on the two sides
of the correspondence and, since then, many checks of the AdS/CFT
duality have been performed.

In an apparently unrelated development~\cite{Gueven:2000ru}, it has
been shown recently that it is possible to extend to supergravity the
limiting procedure described by Penrose in the case of pure gravity
\cite{penrose}.  The nice feature of this limit is that it deforms a
solution of the classical equations of motion into a new universal
wave-like solution.  The Penrose--limit of the $AdS_5 \times S_5$
background in type IIB supergravity corresponds to the plane--wave
solution~\cite{Blau:2002dy},
\beq\label{pwave-back} 
ds^2 = -4 dx^+ dx^- - \mu^2 \sum_{I=1}^8 x_I x^I (dx^+)^2 +
\sum_{I=1}^8 dx_I dx^I ~,~~~ F_{+1234} = F_{+5678} = 2\mu~,
\eeq
which was initially obtained in~\cite{Blau:2001ne} as a new background
preserving all IIB supercharges.  Since, as showed
in~\cite{Metsaev:2001bj,Metsaev:2002re}, IIB string theory in the
above plane--wave background is solvable in the Green-Schwarz
formalism, this limiting procedure turned out to be very interesting
for the AdS/CFT duality. This was first pointed out by Berenstein,
Maldacena and Nastase~\cite{Berenstein:2002jq} who proposed that, on
the Yang--Mills side, this limiting procedure corresponds to focusing
on a subset of the full spectrum of composite operators of ${\cal N} =
4$ Yang--Mills.

The possibility to work with a tractable {\em string} theory showed a
glimpse of the full power of the AdS/CFT duality. In fact, the simple
computation of the mass spectrum in string theory gave an exact
prediction for the conformal dimensions ($\Delta$) of the
corresponding gauge theory operators; that is the $\alpha'$ dependence
of the string masses translates into a continuous function of the
effective coupling interpolating, in the planar limit, between the
weak and the strong--coupling behavior of $\Delta$. This prediction
has been checked on the gauge theory side first at the perturbative
level up to 2--loop order~\cite{Gross:2002su}, and then at all orders
in~\cite{Santambrogio:2002sb}. The behavior of $\Delta$ has been been
studied also at the torus
level~\cite{Kristjansen:2002bb,Constable:2002hw}, where the situation
is more complicated due to a nontrivial mixing of BMN operators.  Of
course, it would be very nice to extend the string theory analysis and
obtain new exact predictions for the ${\cal N} = 4$ Yang--Mills
theory. However, there are still two largely unsolved issues
preventing a straightforward application of the pp--wave/CFT duality
beyond the computation of conformal dimensions.
 
The first problem is that the background (\ref{pwave-back}) gives rise
to a free world--sheet theory only in the light cone gauge. Moreover,
a non--trivial R--R field can be handled, so far, only in the
Green-Schwarz formalism\footnote{In~\cite{Berkovits:2002zv} an
alternative formalism has been applied to the case of pp--wave
background. Even if this looks like a promising step toward a
covariant quantization, the presence of a non--trivial background
gives rise to a complicate world-sheet action, and explicit
computations of string interactions in this framework have not been
done so far.}. In this approach the description of the string
interaction is quite involved already in the flat--space background
and a general form for the tree--level $N$--string amplitude is not
known. However, in the 80's a detailed analysis of this formalism was
carried out~\cite{Green:1982tc,Green:hw} (see also Chap. 11 of
\cite{Green:mn} and references therein) so that the first interesting
amplitudes could be computed. Recently the IIB $3$-string vertex in
the background~(\ref{pwave-back}) has been studied by the authors of
Ref.~\cite{Spradlin:2002ar} extending the flat space analysis
of~\cite{Green:hw}.

A second problem in the pp--wave/CFT duality is represented by the
dictionary between string and gauge theory dynamical computations. The
rule given in~\cite{Gubser:1998bc} for comparing Yang--Mills Green
functions and supergravity tree--level graphs does not seem to be
directly applicable to the gauge theory operators we are interested
in, since the corresponding supergravity excitations are confined far
away from the AdS conformal boundary. It is not even clear whether it
is possible to extract from string theory the complete Yang--Mills
Green functions. Actually, contrary to what happens in the
supergravity limit, we do not expect that this will be possible in
general.  A rule for comparing string interactions and Yang-Mills
results was proposed in~\cite{Constable:2002hw}.  This was  further
investigated and generalized
in~\cite{Kiem:2002xn}--\cite{Spradlin:2002rv}.  The correspondence
discussed in these papers involved a particular class of BMN operators
-- the operators with scalar impurities only.

In this note we consider the extension of the pp--wave/CFT duality to
BMN operators of a different kind and, correspondingly, to different
3-string interactions.  We mainly focus on the string side with the
aim of giving an explanation for the puzzle posed by the Yang--Mills
computation of~\cite{Gursoy:2002yy}, where a BMN operator containing a
derivative impurity ($D_\mu Z$) has been considered explicitly for the
first time. A detailed computation of its conformal dimension is
presented both at the planar and at the torus level, generalizing the
ideas and the results of~\cite{Kristjansen:2002bb,Constable:2002hw} to
this new kind of BMN operators. It turns out that the result
coincides exactly with the one found for a purely scalar BMN
operator. If this result is interpreted via the unitarity argument
presented in~\cite{Constable:2002hw}, one gets information about the
$3$--string tree--level interaction that should be directly compared
with the string cubic vertex of~\cite{Spradlin:2002ar}. However, as
noticed in various points in the
literature~\cite{Constable:2002hw,Lee:2002rm,Spradlin:2002rv,Gursoy:2002yy},
the string interactions seems to be vanishing and would imply a zero
torus-level contribution to the anomalous dimension of the operator
under study.

This is in conflict with the field theory result.
A possible explanation of this mismatch (beyond those suggested
in~\cite{Gursoy:2002yy}) is that the unitarity argument
of~\cite{Constable:2002hw} is incorrect and thus cannot be
reliably used to derive the $3$--point function from the value of the
anomalous dimension. In fact, this possibility has been
confirmed in the very recent literature
based on field theory calculations.

In this note, we  examine the string theory side of the
correspondence. Our main point is that the light--cone quantization
of string theory on the pp--wave background so far considered does not
realize in an explicit way all the symmetries of the background
(\ref{pwave-back}). Thus we propose that some modifications have to be
made and that they affect the structure of the zero--modes in the
interaction Hamiltonian, which is the part responsible for the puzzle
described above.  We argue that it is also possible that the non--zero
mode part of the vertex has to be modified.

The plan of the paper is the following. We start with a discussion of
a discrete $Z_2$ symmetry of the background~(\ref{pwave-back}) and its
realization in the string quantization procedure. We show that
implementing this $Z_2$ symmetry requires a modification in the choice
of vacuum and in the zero--mode structure of the interaction
Hamiltonian. We then focus on the part of the vertex relevant for
computing amplitudes among string states without any fermionic
oscillator. On one hand we show that these modifications do not spoil
the exact matching between gauge and string theory found so far in the
case of scalar BMN operators. On the other hand we are able to
reconcile the $3$--point string computation with the field theory
result of~\cite{Gursoy:2002yy}. In fact it turns out that amplitudes
among string states with only bosonic excitations display the $SO(8)$
symmetry already present in the bosonic part of the
background~(\ref{pwave-back}).

\section{The string interaction and its symmetries}

In this section we will analyze the symmetries of the interaction
Hamiltonian proposed in Ref.~\cite{Spradlin:2002ar}. As it is well
known, symmetries play a crucial role in determining the string
interaction in the Green--Schwarz formalism. Contrary to what happens
in the covariant formalism, one does not have at his disposal a
world-sheet BRST charge that can be used to define vertex
operators. Thus the strategy used in~\cite{Green:1982tc,Green:hw} is
to first look for a string interaction realizing locally on the
world--sheet all the kinematical symmetries of the light--cone
algebra. In the case of flat space, this implies that the string
coordinates are continuous at the interaction point and that the
conjugate momenta are conserved. By applying the same idea to the
pp--wave background, in \cite{Spradlin:2002ar} it was shown that
$|H_3\rangle$ enjoys exactly the same features as in flat space--time,
even if this is not the case for the quadratic part of the
Hamiltonian. Then one has to look at the dynamical part of the
supersymmetry algebra. It turns out that, in order to have a string
interaction respecting the dynamical supersymmetries, one has to add a
particular prefactor term. Also in this part, the analysis
of~\cite{Spradlin:2002ar} follows closely the flat space--time
case~\cite{Green:hw}. However, the bosonic symmetries of the
background~(\ref{pwave-back}) are not those of flat--space and this
suggests to introduce some different choice in the treatment of the
zero--modes, thus yielding a different form of $|H_3\rangle$.

\subsection{The $Z_2$ symmetry and the choice of vacuum}
The point we want to stress is that the presence of a non--trivial
R--R field in the plane--wave background breaks the light--cone
Lorentz symmetry $SO(8)$ down to $SO(4)\times SO(4)\times Z_2$. The
two $SO(4)$ rotate the first and the last four directions among
themselves respectively, while the discrete $Z_2$ symmetry swaps
simultaneously the two groups of four directions
\beq\label{Z2}
Z_2 : \;\; (x_1,x_2,x_3,x_4) ~\leftrightarrow~ (x_5,x_6,x_7,x_8)~.
\eeq
Of course one is free to define the action of $Z_2$ in a different
way, for instance by assigning a minus sign to all the components of
the r.h.s. of~(\ref{Z2}). These different choices, however, are
perfectly equivalent, since they differ just by a rotation in one of
the two $SO(4)$ and generate the same $SO(4)\times SO(4)\times Z_2$
group. Notice that the $Z_2$ transformation above is just a
particular rotation in the full $SO(8)$ group. Indeed, a generic
$SO(8)$ rotation by an angle $\omega_{IJ}$ is $~\exp\left(i
\omega_{IJ} M^{IJ} \right)$, where $M_{IJ}$ are the standard $SO(8)$
Lorentz generators.  For the vector representation $M_{IJ} = i (x_I
\partial_J - x_J \partial_I)$ and the transformation~(\ref{Z2}) is
more easily seen as the composition of two rotations: a first one with
$ \omega_1 = \frac{\pi}{4}(i \sigma_2)\otimes 1_{4\times 4}$ and a
second one with $ \omega_2 = \frac{\pi}{4}(1_{2\times 2} -
\sigma_3)\otimes 1_{2\times 2}\otimes(i \sigma_2)$.

From this observation one can easily derive the explicit action of
$Z_2$ on the spinors. Once a particular realization of the $SO(8)$
$\gamma$--matrices is chosen, it is sufficient to use the appropriate
generators $M^{IJ} = \frac{-i}{4} [\gamma^I,\gamma^J]$ in the above
formula to find the rotation matrix. As noticed in various points in
the literature, computations get simplified if one works with a
specific representation where $\Pi = \gamma^1 \gamma^2 \gamma^3
\gamma^4 = 1_{2\times 2} \otimes\sigma_3 \otimes 1_{4\times 4}$. This
can be realized as follows
\bea\label{gamma}
\gamma^1 = (i \sigma_2)\otimes (i \sigma_2)\otimes (i \sigma_2)\otimes
(i \sigma_2) ~~,& \gamma^2 = (i \sigma_2)\otimes \sigma_1 \otimes (i
\sigma_2)\otimes 1_{2\times 2},  & \non \\
\gamma^3 = (i \sigma_2)\otimes (i \sigma_2) \otimes 1_{2\times
2} \otimes \sigma_1 ~~,& 
\gamma^4 = (i \sigma_2)\otimes (i \sigma_2) \otimes 
 1_{2\times 2}  \otimes \sigma_3, & \\ \non
\gamma^5 = (i \sigma_2)\otimes 1_{2\times 2} \otimes  
\sigma_1 \otimes (i \sigma_2) ~~,& 
\gamma^6 = (i \sigma_2)\otimes \sigma_3 \otimes 
(i \sigma_2) \otimes 1_{2\times 2},  & \\ \non
\gamma^7 = \sigma_1 \otimes 1_{2\times 2} \otimes 
1_{2\times 2} \otimes 1_{2\times 2} ~~,& 
\gamma^8 = (i \sigma_2)\otimes 1_{2\times 2} \otimes 
\sigma_3 \otimes  (i \sigma_2)  &~.
\eea
From the above $\c$'s one can, as usual, construct the $SO(1,9)$ real
and chiral $\C$'s: $\C^0 = (i\sigma_2) \otimes 1_{16\times 16}$ and
$\C^\mu =\sigma_1 \otimes \c^\mu$ for $\mu=1,\ldots,9$, with $\c^9 =
\prod_{I=1}^8 \c^I$. However, we will not need the 10D $\C$'s since
all our spinors are both Majorana-Weyl and satisfy the light--cone
constraint $(\C^0+\C^9)\theta = 0$. This constraint means that, with the
chosen $\c$--representation, only the first eight components of a
spinor are non-vanishing. With the definitions~(\ref{gamma}), one can
easily verify that the $Z_2$ reflection~(\ref{Z2}) simultaneously
exchanges some components of the $8$--dimensional spinor:
\be \label{theta-flip}
\theta^3 \leftrightarrow \theta^4, \quad \mbox{and} \quad \theta^7
\leftrightarrow -\theta^8.
\ee

Notice that the 2D string action in the light--cone
gauge~\cite{Metsaev:2001bj} is $Z_2$ invariant, even if the
combination $\Pi$ appears explicitely. In fact $\Pi$ and $\Pi' =
\gamma^5 \gamma^6 \gamma^7 \gamma^8 = \sigma_3 \otimes\sigma_3 \otimes
1_{4\times 4}$ have exactly the same action on the first eight
components (those relevant for us). The $Z_2$ invariance of the string
action is reflected at the level of the energy spectrum\footnote{See
the detailed analysis of~\cite{Metsaev:2002re} and in particular
tables I and II, apart from the typo in $b_{ij}^\ominus$(6), whose
$SO(4)\times SO(4)$ labels should be $(1,-1)\times (0,0)$.}.

It is then natural to require that the $Z_2$ symmetry~(\ref{Z2}) is
preserved by the interaction terms of the Hamiltonian. At first sight
this seems obvious. In fact the construction in~\cite{Spradlin:2002ar}
parallels the one of~\cite{Green:hw} and thus one may think to have a
3-string vertex which is invariant under the full $SO(8)$. However
this is not the case. In the notations of~\cite{Spradlin:2002ar}
and~\cite{Spradlin:2002rv}, the 3-string vertex reads:
\beq\label{vs}
|H_3\rangle =\left[ K_I \widetilde{K}_J v^{IJ}(\Lambda)~ E_a\; E'_b\;
E^0_b \right] |0\rangle_1 \otimes |0\rangle_2 \otimes |0\rangle_3~. 
\eeq
For our purposes, we have separated the contributions of the fermion
non-zero modes $E'_b$ and of the fermion zero modes $E^0_b$. We will
argue below that~\eq{vs} does not respect the $Z_2$ symmetry of the
background and that both $|0\rangle_1 \otimes |0\rangle_2 \otimes
|0\rangle_3$ and $E^0_b$ have to be modified.

The point is that the Lorentz structure of the operator inside the
square brackets is very similar to the flat--space vertex, the only
novelty being the presence of the $Z_2$-invariant combination
$\Pi$. Thus the expression in the brackets commutes with the $Z_2$
generator. However one has still to define the action of~(\ref{Z2}) on
the ``vacuum'' state $| 0\rangle$ defined by
\be
a_n |0\rangle =0, \forall n, 
\quad b_n |0\rangle =0, n \neq 0, \quad \th_0 |0\rangle =0.
\ee
In order to have a $Z_2$ preserving interaction it is natural to
define
\beq\label{oc}
Z_2 | 0\rangle = | 0\rangle~.
\eeq
This choice also guarantees that in the limit $\mu\to 0$, one smoothly
recovers the flat space theory, where the vacuum is a $SO(8)$
scalar. However, the innocent looking choice~(\ref{oc}) is quite
strange. In fact, as shown in~\cite{Metsaev:2002re}, $|0\rangle$
preserves the full $SO(8)$ symmetry, since it is defined as
$\theta^a_0 |0\rangle= 0$, but it is not the state of minimal
light--cone energy $|\rv \rangle$. Its energy is given by $4 \mu$.
The vacuum state $|\rv \rangle$ with zero energy is defined by
\be
a_n |\rv \rangle = b_n |\rv \rangle =0 \quad \forall n.
\ee
The definition of the ``zero--mode'' oscillators $b_0^a$ in terms of
the $\theta_0^a$ is given in Eq.~(\ref{t0b}). This implies that $|\rv
\rangle$ is related to $|0\rangle$ as follows (for example, for
positive $p^+$):
\beq\label{0v}
|0\rangle = \theta^5_0\,\theta^6_0\,\theta^7_0\,\theta^8_0
|\rv \rangle~ .
\eeq
The relations~\eq{theta-flip}, \eq{oc} and~\eq{0v} imply that $|\rv
\rangle$ is odd under $Z_2$, since we found that under this
transformation $\theta^7$ and $\theta^8$ are exchanged.  We note that
the minus sign arising from the action of $Z_2$ on the product of
$\th$'s above does not depend on the particular realization of the
$\c$-matrices chosen in~(\ref{gamma}). The specific form of the action
may change, but the minus sign is always present. This is consistent
with the supergravity analysis of~\cite{Metsaev:2002re} where the
polarizations of the string ``massless'' states are mapped into the
various supergravity states. One can check that the lowest energy
field $h$ is odd under $Z_2$. This flip of sign in the {\it field}
matches with the fact that the product of the four $\theta$'s
in~\eq{0v} represents the polarization of the field.

Thus $|\rv \rangle$ and $|0\rangle$ cannot have the same
$Z_2$--parity. With the choice~\eq{oc} $|0\rangle$ is $Z_2$-invariant.
However this assignment is puzzling both on the string and the gauge
theory side.  On the string side, defining $|0\rangle$ to be a scalar
is natural only in flat--space where all the supergravity modes are
degenerate in energy and $|0\rangle$ plays a special role, being
$SO(8)$ invariant. In the pp--wave background, however, (\ref{oc}) is
not natural because the zero--modes are not degenerate in energy; it
is the state $|\rv \rangle$ that plays a special role because it
represents the real vacuum of the theory ({\em i.e.} the state of
minimal energy and supersymmetry preserving). The conventional choice
in string quantization is to define the vacuum to be invariant under
all the symmetries of the background (including the discrete
symmetries, present, for instance, in orbifold
compactifications). This choice is also supported by the analysis of
different, but closely related, setups\footnote{ For instance, the
pp--waves background supported by a NS--NS flux were reconsidered
in~\cite{Kiritsis:2002kz} with the goal of studying the holographic
properties of these backgrounds. These pp--waves can be analyzed by
means of CFT techniques in the RNS quantization, and it turns out that
the ground state is always invariant under all discrete symmetries of
the background. Moreover, it is interesting to note that the Matrix
String Theory analysis of certain pp--wave backgrounds also yields a
symmetry--preserving interaction~\cite{Bonelli:2002mb}.}.

The assignment~(\ref{oc}) is also puzzling from the point of view of
the string/gauge--theory correspondence. In fact, $|\rv \rangle$
corresponds to the operator $O_{{\rm vac}}^{J} \sim {\rm tr}\, Z^J$
which is naturally defined to be $Z_2$ invariant, since it does not
have any Lorentz index along the directions where the $Z_2$ action is
non--trivial.

Thus, there are two reasons why one should change the quantum number
assignment of $| \rv \rangle$ and declare it to be a $SO(4)\times
SO(4)\times Z_2$ scalar. The first reason is to construct a string
interaction explicitly realizing the $SO(4)\times SO(4)\times Z_2$. The
second reason is to keep a close relation with the gauge--theory
correspondence. In particular,
\beq\label{ec}
Z_2 | \rv \rangle = | \rv \rangle~~~~\Leftrightarrow~~~~
Z_2 | 0\rangle = - | 0\rangle~.
\eeq
This change however is not without consequences since it implies that
the $3$--string interaction~(\ref{vs}) does not preserve the $Z_2$
invariance of the background~(\ref{pwave-back}).  Notice that, since
$Z_2^2=1$, (\ref{oc}) and (\ref{ec}) are the only 2 possible
assignments, if one insists that $|\rv \rangle$ is an eigenstate of
$Z_2$.

In summary, the more desirable choice~(\ref{ec}) is possible only if
the form of the interacting Hamiltonian~\eq{vs} is modified too.
 
\subsection{The string interaction $H_3$}

The existing form~\eq{vs} for $|H_3\rangle$ and its behavior under
$Z_2$ are responsible for a few puzzling features noticed in the
literature. The authors of~\cite{Constable:2002hw} noticed that
$3$-string amplitudes involving only states dual to scalar BMN
operators have a relative minus sign with respect to other amplitudes
involving only operators with derivative impurities.  Even more
puzzling, it seems that string theory predicts a vanishing $3$-point
interaction when the incoming state is dual to an operator with one
scalar and one derivative impurity.  These properties of the string
amplitudes have been checked by explicit
computations~\cite{Kiem:2002xn,Lee:2002rm,Spradlin:2002rv}, and follow
from the vacuum choice~(\ref{vs}) and the $\c$-matrix relations
\beq\label{sr}
\sum_{K=1}^8 \c^{i K}_{[ab}\c^{jK}_{cd]} =\delta^{ij}
\epsilon_{abcd}~, \qquad
\sum_{K=1}^8 \c^{i' K}_{[ab}\c^{j'K}_{cd]} =  -\delta^{i'j'}
\epsilon_{abcd}~, \qquad
\sum_{K=1}^8 \c^{i K}_{[ab} \c^{j'K}_{cd]} = 0~,
\eeq
where the spinor indices $a,b,\ldots$ are restricted to be in the
positive $\Pi$--chirality ({\em i.e.} $a=1,\ldots,4$) or in the
negative $\Pi$--chirality ({\em i.e.} $a=5,\ldots,8$) subspace, while
the vector indices $i,j$ run from $1$ to $4$, $i',j'$ from $5$ to $8$
and $K$ runs from $1$ to $8$. Eq.~(\ref{sr}) is derived by a direct
computation from the $\c$--matrix realization~(\ref{gamma}) and the
relative minus sign between the first two relations is responsible for
the puzzling features described above. However, in order to see the
effects of the minus sign in Eq.~(\ref{sr}), one does not have to go
through the $\c$--matrix algebra. It is sufficient to use the
transformation rules under $Z_2$ of the $3$--string amplitudes.  Let
us define the string amplitude
\be \label{Aij} 
A^{IJ}  :=  \left(\langle \rv,p_3^+| \a_{n (3)}^{I}\a_{-n (3)}^{J} \otimes
\langle \rv,p_1^+| \a_{m (1)}^{I}\a_{-m (1)}^{J} \otimes
\langle \rv,p_2^+|\; \right)\; |H_3 \rangle,  
\ee 
and consider the string amplitudes $A^{ij}, A^{i'j'}$. The
oscillators\footnote{As already said we follow the conventions
of~\cite{Spradlin:2002ar} or~\cite{Spradlin:2002rv} and denote with
$\alpha_n$ the BMN string oscillators and with $a_n$ the oscillators
usually employed in the string vertex. The relations are $\a_n =
\frac{1}{\sqrt{2}} (a_{|n|} - i\; {\rm sign}(n) a_{-|n|} )$, for $n
\neq 0$ and $\a_0= a_0$.} inserted in $A^{i'j'}$ are just the $Z_2$
images of those inserted in $A^{ij}$. Moreover the operatorial content
of the interaction~(\ref{vs}) is $Z_2$ invariant. Thus one can insert
$Z_2^2 = 1$ in the amplitude $A^{ij}$ and relate it to $A^{i'j'}$ with
a coefficient of proportionality of $(-1)^3$. In fact, the states
$|0\rangle$ and $|\rv \rangle$ cannot be even at the same time and the
factor of $(-1)$ comes either from the action of $Z_2$ on the external
states or from the action on the kets in $|H_3\rangle$, according to
the $Z_2$--parity chosen for $|\rv \rangle$. Thus one is led to the
conclusion that $ A^{ij} = - A^{i'j'}$. A similar argument, where one
uses in addition the $SO(4)\times SO(4)$ invariance and the fact that
the exchange $(n,m)\to (-n,-m)$ leaves invariant the oscillator
contribution, implies that the mixed amplitudes
$A^{ij'}=0$~\footnote{We emphasize that these relations for $A^{ij},
A^{i'j'}$ and $A{ij'}$ follow immediately from the fact that $|0
\rangle$ is used in the construction of the vertex~\eq{vs}, while
the perturbative string states are built on the true vacuum
$|\rm{v}\rangle$. Thus they are independent of the choice of $Z_2$
parity assignment given to $|0 \rangle$ and
$|\rm{v}\rangle$.}. Through more direct computations, these features
of the vertex~\eq{vs} were also noted
in~\cite{Lee:2002rm,Spradlin:2002rv,Gursoy:2002yy}.

It is clear that, in order to obtain a different result for the
amplitudes, it is not sufficient to switch from choice~(\ref{oc})
to~(\ref{ec}) keeping the operator part of $|H_3\rangle$ unchanged.
So we will modify the operator part of $|H_3\rangle$ in order to
construct a string interaction where both the 3-point vertex and the
vacuum are $Z_2$-invariant (as it happens in flat--space with the
$SO(8)$ invariance).

For this purpose, we just need to focus on the fermionic
zero--modes. This part is constrained by the requirements that the
string coordinates $\th$ are continuous and that the conjugate
momentum $\l$ is conserved. This means that we should find a state
$|\delta\rangle$ in the product of the three Hilbert spaces of the
external strings satisfying simultaneously
\beq\label{fcon}
\sum_{\rr=1}^3 \lambda_{0 (\rr)}^a |\delta\rangle = 0 ~~,\qquad
\sum_{\rr=1}^3 p^+_{\rr} \th_{0 (\rr)}^a |\delta\rangle = 0 ~,
\eeq
where $\lambda_{(r)}$ are the conjugate momenta of the fermions
$\theta_{(r)}$. The solution of~\cite{Spradlin:2002ar} is basically
unique if one wants to keep the $SO(8)$ invariance.  However, in the
pp--wave background the symmetry is reduced and it is the smaller
symmetry $SO(4)\times SO(4)\times Z_2$ that has to be preserved.  This
can be achieved by choosing
\beq\label{newdelta}
|\delta\rangle = \prod_{a=1}^8
\left(\sum_{\rr=1}^3 \lambda_{0(\rr)}^a\right) 
\prod_{a=1}^8 \left(\sum_{\rr=1}^3 p^+_{\rr} \th_{0 (\rr)}^a\right)
|\rv,p_1^+\rangle \otimes |\rv,p_2^+\rangle \otimes
|\rv,p_3^+\rangle~.
\eeq 
Here the fermionic delta function $|\delta\rangle$ satisfies the
constraints~(\ref{fcon}). In fact, because of momentum conservation
$\sum_{\rr=1}^3 p^+_{\rr} = 0$, the operators $\sum_{\rr=1}^3
\lambda_{0 (\rr)}^a$, $\sum_{\rr=1}^3 p^+_{\rr} \th_{0 (\rr)}^b$
anti-commute for all $a,b$. Therefore in checking~(\ref{fcon}), one
always encounters the square of a fermion oscillator and~\eq{fcon} are
satisfied.  Notice that in Eq.~(\ref{newdelta}) it is crucial to use
the $SO(8)$ breaking vacuum $|\rv \rangle$. If the ground state
$|0\rangle$ had been used in the r.h.s. of~(\ref{newdelta}), one would
have found a trivially vanishing result. Thus our proposal is to use
this new solution to the fermionic constraints~(\ref{fcon}) and
replace the combination $E^0_b |0\rangle_1 \otimes |0\rangle_2 \otimes
|0\rangle_3 $ in the vertex (\ref{vs}) with the fermionic
delta-function defined in~(\ref{newdelta}):
\beq\label{newH3}
|H_3\rangle^{{\rm new}}
 =\left[ K_I \widetilde{K}_J v^{IJ}~ E_a\; E'_b\ \right] 
|\delta \rangle~. 
\eeq
However the explicit form of $K_I$, $\widetilde{K}_J$ and $v^{IJ}$ is
likely to be different from the one of flat--space. For the purposes
of this letter we will only need to assume that $v^{IJ}$ contains a
constant term proportional to $\delta^{IJ}$.

\subsection{Consequences on the bosonic amplitudes}

An immediate advantage of the above modification is that now the
zero-mode structure of the interaction Hamiltonian is $Z_2$--even with
the natural choice~(\ref{ec}). Thus the argument yielding $A^{ij}
=-A^{i'j'}$ cannot be applied if~(\ref{newdelta}) is used and one can
hope to avoid the puzzling features deriving from the
interaction~(\ref{vs}). Now we will prove that this is indeed the
case.

Let us focus on string amplitudes involving external states built with
only bosonic oscillators acting on the true vacuum $|\rv \rangle$. In
this case we will not need the precise form of the prefactor
$v^{IJ}$. In fact, as noticed in~\cite{Green:hw} the prefactor can
only contain creation oscillators. The lowering modes are irrelevant
since they can read the vacuum structure of the vertex. The creation
modes are defined with respect to the vacuum chosen in the interaction
Hamiltonian. Since we write the string vertex in terms of the new
zero--mode structure~(\ref{newdelta}), the prefactor can only contain
$a^\dagger$ and $b^\dagger$ oscillators. If we use external states
with no fermionic oscillators, all the terms in the prefactor
containing $b^\dagger$'s will not contribute to the amplitude, since
they can act directly on the external states. This means that the only
term of the prefactor matrix $v^{IJ}$ we will need is the constant
part $v^{IJ} = \delta^{IJ}$. This is conflict with the form $v^{IJ} =
\pm\, \delta^{IJ} $, $+$ for $I,J=1,2,3,4$, $-$ for $I,J=5,6,7,8$, as
obtained in \cite{Lee:2002rm,Spradlin:2002rv}, which would give the
following string amplitudes
\be
A^{ij} = - A^{i'j'}, \quad A^{ij'} =0, \quad \mbox{if one uses 
$|H_3\rangle$ in  Eq.~\eq{vs}}.
\ee
On the other hand, it is clear that, due to the presence of the
$SO(8)$ invariant $v^{IJ} =\d^{IJ}$, the string vertex~\eq{newH3}
leads to
\be
A^{ij} = A^{i'j'} = A^{ij'}, \quad \mbox{if one uses 
$|H_3\rangle^{{\rm new}}$ 
in Eq.~\eq{newH3}}.
\ee
Notice that these amplitudes are nonzero. To see this, let us denote
the fermionic part common to all of them by
\be
F:= {}_1\langle \rv| \otimes {}_2\langle \rv|\otimes {}_3\langle \rv|  
\prod_{a=1}^8 \left(\sum_{\rr=1}^3 \lambda_{0(\rr)}^a\right)
\prod_{a=1}^8 \left(\sum_{\rr=1}^3 p^+_{\rr} \th_{0 (\rr)}^a\right) 
| \rv \rangle_1 \otimes | \rv \rangle_2 \otimes | \rv \rangle_3.
\ee
Without loss of generality, let us take $p^+_1, p^+_2 >0, p^+_3
<0$. In terms of string oscillators $b$'s, we have (apart from
numerical factors)
\be\label{t0b}
\th_{0(\rr)} \sim \frac{1}{\sqrt{p^+_{\rr}}}
\pmatrix{ b_{0(\rr)}\cr b^\dagger_{0(\rr)}}, 
\quad \mbox{\rr=1,2};\qquad  
\th_{0(3)} \sim  \frac{1}{\sqrt{|p^+_{3}|}}
\pmatrix{ b^\dagger_{0(3)}\cr b_{0(3)}}, 
\ee
thus
\bea
&&\prod_{a=1}^8 \sum_{\rr=1}^3 p^+_{\rr} \th_{0 (\rr)}^a 
| \rv \rangle_1 \otimes | \rv \rangle_2 \otimes | \rv \rangle_3 \nonumber\\
&&= (p^+_3)^2 \cdot {b^{1\;\dagger}_{0(3)}} \cdots 
{b^{4\;\dagger}_{0(3)}}
(\sum_{\rr=1}^2 \sqrt{p^+_{\rr}} b^{5\;\dagger}_{0(\rr)}) \cdots 
(\sum_{\rr=1}^2 \sqrt{p^+_{\rr}} b^{8\;\dagger}_{0(\rr)})
| \rv \rangle_1 \otimes | \rv \rangle_2 \otimes | \rv \rangle_3 .
\eea
The eight fermion creators have to be annihilated by selecting the
corresponding eight annihilators from the factor $\prod_{a=1}^8
\sum_{\rr=1}^3 \lambda_{0(\rr)}^a$. From the expression in terms of
oscillators
\be\label{l0b}
\lambda_{0(\rr)} \sim \sqrt{p^+_{\rr}} 
\pmatrix{ b^\dagger_{0(\rr)}\cr b_{0(\rr)}}, 
\quad \mbox{\rr=1,2};\qquad  
\lambda_{0(3)} \sim \sqrt{|p^+_{3}|} \pmatrix{ b_{0(3)}\cr b^\dagger_{0(3)}},
\ee
and $\{ \sum_{\rr=1}^2 \sqrt{p^+_{\rr}} b^{a\;\dagger}_{0(\rr)},
\sum_{\rr=1}^2 \sqrt{p^+_{\rr}} b^{b}_{0(\rr)} \} = -p^+_3 \d^{ab} $,
it is easy to obtain
\be
F = (p^+_3)^8.
\ee
%

\section{Discussion} 

In this note we proposed a modification in the treatment of the
fermionic zero--modes which explicitly satisfies all the symmetries of
the plane--wave background.  We also found that the new form of the
string vertex gives results in direct agreement with the expected
$SO(8)$ symmetry of the bosonic excitations\footnote{Notice, in fact,
that the metric in~(\ref{pwave-back}) is $SO(8)$ symmetric and the
breaking of this group is just due to a term in the {\em fermionic}
Lagrangian coming from the R-R form.}. Moreover, the interaction
vertex discussed here matches in a simple way, from the string point
of view, the explicit Yang--Mills computation of~\cite{Gursoy:2002yy}.
However, one may think that even a small modification in the string
vertex could ruin all the subtle cancellations necessary to have a
consistent realization of the supersymmetry algebra on the interaction
Hamiltonian. One may also worry about the assignment~(\ref{ec}),
because it is not what is done in flat--space. Let us first make some
comments on this second issue.

The change of the definition of the fermionic vacuum in the pp--wave
background is easy to understand. In fact an analogous change in the
definition of the bosonic vacuum has already been
performed~\cite{Spradlin:2002ar}. For strings in flat--space, one
considers the eigenstates of the position or momentum operator as
possible vacua. In particular, it has been known for a long time that
the $3$--string interaction contains the following vacuum
structure~\cite{Sciuto:1969vz}
\beq\label{fv}
|0\rangle_{\rm f.s.} = \prod_{\rr=1}^3 | x_{0 (\rr)}^I =0;0_a\rangle =
 \prod_{\rr=1}^3 \int d p^I_{(\rr)} \; |p^I_{(\rr)};0_a\rangle ~.
\eeq
However, when $\mu$ is switched on in the world-sheet Lagrangian both
the fermionic and the bosonic coordinates acquire a potential term of
the harmonic oscillator type.  Thus one cannot use the eigenstates of
the position or momentum operator any more and has to change the
definition of the vacuum. As usual the combination $a^i =
\frac{1}{\sqrt{2 m}} (p^i + i m x_0^i)$ is introduced and the vacuum
is defined as $a^i |0_{\rm a}\rangle=0$ (here $m$ is, of course,
proportional to $\mu$). This vacuum also
appears~\cite{Spradlin:2002ar} in the $3$--string interaction, instead
of the one~\eq{fv} typical of flat space.  This is precisely what is
usually done in quantum mechanics when passing from a free particle to
the case of a harmonic oscillator. What we are claiming is that also
on the fermionic side, a different vacuum $| \rv \rangle$ has to be
chosen once a nonzero $\mu$ is turned on.  We also claim that the
vacuum state $| \rv \rangle$ has to be a scalar in order to realize
explicitly the $SO(4)\times SO(4)\times Z_2$ symmetry of the
background.

The new realization of the fermionic constraints~\eq{newdelta} may
have consequences also on the general structure of the vertex and, in
particular, of the prefactor. The kinematical part of the fermionic
vertex can in principle be constructed exploiting the same idea used
for the zero--modes. In fact, one can just multiply $|\delta\rangle$
by all the other modes of the two fermionic constraints (momentum
conservation and coordinate continuity) and obtain an expression for
$E'_b$ in Eq.~(\ref{newH3}) satisfying all the kinematical
constraints. On the other hand, the explicit form of the prefactor is
related to the realization of the dynamical generators and is more
subtle. However, there is also a technical reason to suspect that the
functional form of the fermionic vertex has to be changed with respect
to the flat-space case. In fact, following the computation
of~\cite{Green:hw}, one sees that the realization of the supersymmetry
algebra on $|H_3\rangle$ requires the identity $(\th_{0 (1)}^a -
\th_{0 (2)}^a) |V\rangle = 0$ (here with $|V \rangle$ we indicate the
ket state in the interaction Hamiltonian enforcing the string
coordinate continuity and momentum conservation). In flat space this
identity simply follows from the fact that the zero--modes
$\th_{(0)}^a$ are all destruction operators, since the vacuum
$|0\rangle$ was used. Notice that this choice is basically forced by
the requirement of $SO(8)$ invariance. On the pp--wave background, we
have to use the real vacuum $| \rv \rangle$ to construct the vertex
$|V\rangle$, so one may have to change this step of the derivation.
Thus the presence of $|\rv \rangle$ may affect the realization of the
supersymmetry algebra on $|H_3\rangle$, and the form of the prefactor
in the complete vertex may be different from that appearing in the
flat--space expression.

Let us conclude by noticing that the vertex in Eq.~(\ref{vs}) yields a
third puzzling prediction on the Yang--Mills side. In fact, beyond
$A^{ij} = - A^{i'j'}$ and $A^{ij'} = 0$ already discussed, it turns
out that all the amplitudes with external states dual to operators
with just fermionic impurities should vanish. This again looks strange
from the field theory point of view.  This zero is not related to
fermion zero--mode counting as the other two problems, but is instead
a consequence of the full form of the interaction~\eq{vs} where the
vertex is at least quadratic in the bosonic oscillators.  We have
presented evidence in the above that the actual form of the functional
prefactor may be different from the one of the flat--space
vertex. Thus in order to see what are the predictions of the parity
conserving interaction~(\ref{newH3}), we need to work out the exact
form of the new $K^I$, $\tilde{K}^I$, $v^{IJ}$ and $\Lambda^a$. 

\vspace{.5cm} {\bf Note added:} The zero--mode
structure~(\ref{newdelta}) proposed in this paper was used to build a
full kinematical vertex in \cite{Chu:2002wj}.  Two supersymmetric
completions are possible for this vertex and have indeed been
discussed in subsequent literature.  One was obtained in
\cite{Pankiewicz:2003kj} by requiring the continuity of the flat space
limit $\mu\to 0$, which implies assigning an {\it odd} $Z_2$ parity
both to the string vacuum $|v\rangle$ and the prefactor. In
\cite{Pankiewicz:2003ap} the resulting vertex was shown to be
equivalent to that proposed in
\cite{Spradlin:2002ar,Spradlin:2002rv,Pankiewicz:2002gs,Pankiewicz:2002tg}.
In \cite{DiVecchia:2003yp} an alternative solution is put forward,
where the choice proposed in this paper of an {\it even} $Z_2$ parity
for the string vacuum is mantained, and this symmetry is therefore
realized explicitly ({\it i.e.} both the interaction and the vacuum
state are $Z_2$--invariant at the same time). In this approach, one
gives up the smoothness of the flat space $\mu\to 0$ limit. In fact,
this second solution follows the behaviour of supergravity in
$AdS_5\times S^5$ more closely. The idea is that, since the PP--wave
background can be seen as an approximate description of $AdS_5 \times
S^5$, even for small curvatures, the 3--state interaction has to be
compared with the results in $AdS_5 \times S^5$ rather than with the
results of flat--space.

Concerning the comparison with the field theory description, two
different approaches have been proposed, \cite{Constable:2002hw} and
\cite{Gross:2002mh} (see~\cite{report} for an updated discussion).
This paper is placed in the framework of \cite{Constable:2002hw},
where string theory amplitudes
are compared with field theory correlators. 
This proposal was explicitely checked
 in~\cite{Kiem:2002xn,Huang:2002wf} for scalar BMN operators and these
 computations were subsequently extended also to BMN operators
 containing vector and fermion impurities \cite{DiVecchia:2003yp}.
Finally, we would like to remark that a more general motivation for
the proposal of \cite{Constable:2002hw} was provided in
\cite{Dobashi:2002ar,Yoneya:2003mu} by considering the Penrose limit
of the AdS/CFT bulk--to--boundary formula of
\cite{Gubser:1998bc}. This approach was extended in
\cite{Huang:2002yt}, where non--planar corrections to the gauge theory
correlators are reproduced from the string side.

\section*{Acknowledgments}
CSC thanks Jan Plefka, Matthias Staudacher, Volker Schomerus, Stefan
Theisen for helpful discussion. We would like to thank Alberto
Zaffaroni for a very important remark. We acknowledge grants from
ESPRC, Nuffield foundation, PPRAC of UK, NSC of Taiwan and INFN of
Italy, and the support of the European Union under RTN contract
HPRN-CT-2000-00131.

\end{document}